\documentclass[twocolumn,prb,showpacs,multicol,amsmath,amssymb]{revtex4}
\usepackage[dvips]{graphicx}
\usepackage{graphicx}
\usepackage{dcolumn}
\usepackage{bm}
\usepackage{graphics}
\usepackage{epsfig,color}

\newcommand{\be}{\begin{equation}}
\newcommand{\ee}{\end{equation}}
\newcommand{\bea}{\begin{eqnarray}}
\newcommand{\eea}{\end{eqnarray}}

\begin{document}
\title{The Effect of FM Inter-ladder coupling in spin-1/2 AFM two-Leg Ladders in the presence of a magnetic field: Quantum Monte Carlo Study}

\author{J. Jahangiri, H. Hadipour, S. Mahdavifar, and S. Farjami}
\affiliation{Department of Physics, University of Guilan, 41335-1914, Rasht, Iran}

\date{\today}


\begin{abstract}
  We study the effect of inter-ladder ferromagnetic (FM) coupling in spin-1/2 two-leg ladders with antiferromagnetic (AFM) legs and rungs interactions using the stochastic series expansion
    quantum Monte Carlo. We have compared the results
      with the experimental measurement on Sr$_{14-x}$Ca$_{x}$Cu$_{24}$O$_{41}$ cuprate which is the candidate for spin-1/2 AFM two-leg
       ladders with FM inter-ladder interaction. A remarkable asymptotic behavior of susceptibility is observed at very low temperature.
In the absence of the magnetic field, thermodynamic behavior of  an individual spin-1/2 two-leg ladder is similar to coupled one up to $-0.5J_{leg}$ interaction. But, the gaped phase is not clear in the $J_{in}=-0.2 J_{leg}$ of FM coupled two-leg ladder even at low magnetic fields, which shows that the inter-ladder FM interaction can induce a new quantum ordered gapless phase at zero temperature.

\end{abstract}


\pacs{75.10.Jm; 75.10.Pq}

\maketitle


\section{Introduction}\label{sec1}

The study of the antiferromagnetic (AFM) spin-1/2 ladder systems has led to
 a huge growth of interest due to crossover from chains to square
  lattices\cite{Dagotto}.  The formation of spin singlets located on each
   rung opens the spin gap in the ground state of an even-leg ladders
    which is called the gaped spin liquid \cite{Rieira}.  This kind of two-leg
        ladders are found in SrCu$_{2}$O$_{3}$ \cite{Azuma} and recently in some copper based
   materials like (C$_{5}$H$_{12}$N)$_{2}$CuBr$_{4}$\cite{Ruegg,Watson},
   (C$_{7}$H$_{10}$N)$_{2}$CuBr$_{4}$\cite{Hong} , TlCuCl$_{3}$ \cite{Oosawa} and
Cu$_{2}$(C$_{5}$H$_{12}$N$_{2}$)$_{2}$Cl$_{4}$\cite{Giamarchi}, vanadate compound MgV$_{2}$O$_{5}$ \cite{Millet,Elfimov,Korotin}, and cuprate superconductor Sr$_{14-x}$Ca$_{x}$Cu$_{24}$O$_{41}$ \cite{McCarron,Siegrist,Roth,McElfresh}.
   The interest in the study of spin two-leg ladders has been strengthen by the realization that
superconductivity occurs in the Sr$_{14-x}$Ca$_{x}$Cu$_{24}$O$_{41}$ cuprate systems \cite{Uehara}.

 The effect of an external magnetic field ($h$) on the spin-1/2 two-leg ladder systems is well-known.  Generally, at low magnetic fields ($h < h_{c_1}$),  there is a spin
liquid phase (a gapped phase) at low temperature \cite{Dagotto}. Both magnetic
susceptibility and the magnetization go up first with cooling, then
decay exponentially to zero at low temperatures. Also, the specific
heat has a single peak at low temperature due to transition from disordered phase to spin singlet gapped phase\cite{Ruegg,Wang}.
 The Tomonaga-Luttinger liquid (TLL) gapless phase is found in $h_{c_1} < h <
h_{c_2}$ regime at low temperatures\cite{Ruegg,Wang}.
 One of the spin liquid ($h <
(h_{c_1} + h_{c_2})/2$) or spin polarized ($h > (h_{c_1} +
h_{c_2})/2$) phases at higher temperature is expected. The
thermodynamic properties like magnetization and the susceptibility
have a finite value at low temperature which show the vanishing of
the energy gap in the TLL phase. Specific heat shows a second peak
and goes down linearly with lowering temperature in the TLL regime \cite{Ruegg}.
At high
magnetic field ($h > h_{c_2}$), lowering the temperature causes a
spin polarized phase. Magnetization goes up and saturates
exponentially at low temperature.  The second low temperature peak
disappears in the specific heat by applying of the strong magnetic field.

A ferromagnetic (FM) inter-ladderin two-leg spin ladders like  interaction exists in two-leg spin ladders like SrCu$_{2}$O$_{3}$, cuprate superconductor Sr$_{14-x}$Ca$_{x}$Cu$_{24}$O$_{41}$ as well as other compound such as  vanadate compound MgV$_{2}$O$_{5}$ \cite{Millet,Elfimov,Korotin}. In the case of SrCu$_{2}$O$_{3}$, the inter-ladder interaction is accurately ferromagnetic.  The origin of FM coupling comes from frustration between two leg ladders in the case of Sr$_{14-x}$Ca$_{x}$Cu$_{24}$O$_{41}$ and MgV$_{2}$O$_{5}$. For example, the triangular scheme of Cu ions on adjacent ladders coupled in this way leads to frustration in Sr$_{14-x}$Ca$_{x}$Cu$_{24}$O$_{41}$ \cite{McCarron}. The structure of these compounds consists of trellis layers.
 Aharony and coworkers indicated that one can replace  the frustrated AF inter-ladder couplings of the trellis lattice by an effective FM inter-ladder interaction in a square lattice \cite{Aharony}.    Inelastic neutron scattering yields the exchange interaction of about $J_{rung}$=800 K for  Sr$_{14}$Cu$_{24}$O$_{41}$ compound \cite{Eccleston}.
Different values of $J_{rung}$/$J_{leg}$= 0.5 and $J_{rung}$/$J_{leg}$= 1 are found by INS measurements and Raman spectroscopy respectively \cite{Sugai,Gozar}.
 Even with considering of frustration, $J_{in}$ is one order of magnitude smaller than the intra-ladder interaction \cite{Gopalan}.  Because the small value of $J_{in}$, one may consider the system as isolated ladders.
Experimental results show a broad peak at 80 K with an upturn below 20 K in the temperature dependence of the magnetic susceptibility \cite{Hiroi}. The value of the spin gap is not sensitive to inter-ladder coupling \cite{Gopalan}, also the magnetic susceptibility is modified at low temperature \cite{Gelle}.

 From theoretical point of view, the effect of the inter-ladder FM coupling in spin-1/2 two-leg ladders with AFM legs and rungs is much less studied. Miyahara and coworkers \cite{Miyahara} performed quantum monte carlo (QMC) simulations on the FM coupled two-leg ladder (trellis layer) system with $J_{in}/J_{leg}=-0.1, -0.2$, and $-0.5$ in absence of the magnetic field. The intra-ladder interactions were considered to be both $J_{rung}$/$J_{leg}$=0.5 and $J_{leg}$/$J_{leg}$=1.0.
The results show that inter-ladder interaction (due to frustration) can not change the behavior of magnetic susceptibility curve even at high value of $J_{in}$.

To the best knowledge of the authors, there is no simulated data on the coupled FM two-leg ladders in presence of the magnetic field. Such an interesting
properties of spin ladder systems call for the investigation of the role of FM inter-ladder interaction in two-leg ladder systems and its evolution upon the increasing of the magnetic field.
So, to study the thermodynamic properties of trellis layer systems in the magnetic fields, here we implement the method
developed stochastic series expansion (SSE) QMC\cite{Sandvik, Syljuasen} for a ladder system with FM interaction between neighboring ladders. We have found good agreement between the experiment
and our QMC calculation for magnetization,  magnetic susceptibility, and the specific heat.


\section{Results and discussion}\label{sec3}

To study the effect of FM inter-ladder exchange interaction on the spin-1/2 two-leg ladder
 systems, we first perform QMC to obtain thermodynamic
  properties, in particular magnetic susceptibility, the magnetization and the specific heat, for an isolated two-leg ladder using experimental parameters obtained by Refs. \cite{Watson} and \cite{Ruegg}. The QMC simulation is performed for ladders with the size of $20\times2$ under the periodic boundary condition with the maximum $1000000$ equilibration sweeps and $2000000$ measurement steps.  Later, we will point out the
    results for some ladders which are coupled ferromagnetically. To compare the calculation with the experimental results, we have performed QMC for isolated spin-1/2 two-leg ladder in the strongly coupled range of interaction $J_{rung}\gg J_{leg}$. But, calculation for FM coupled spin-1/2 two-leg ladder has been done with the exchange interaction $J_{rung}$=$J_{leg}$. We will show that the spin singlet state (gaped phase) exists even in the $J_{rung}$=$J_{leg}$ which is consistent with the results of Dagotto et al. paper \cite{Dagotto}. To compare better the FM coupled ladders results with isolated one, we consider $J_{leg}$=$J_{rung}$=3.9 K for coupled ladders system which is two order of magnitude less than experiment. So, both temperature range of Curie-Weiss behavior and magnetic critical fields in our QMC simulation for the FM coupled ladders are lower than experimental results of Sr$_{14-x}$Ca$_{x}$Cu$_{24}$O$_{41}$.


             \begin{figure}[t]
\centerline{\psfig{file=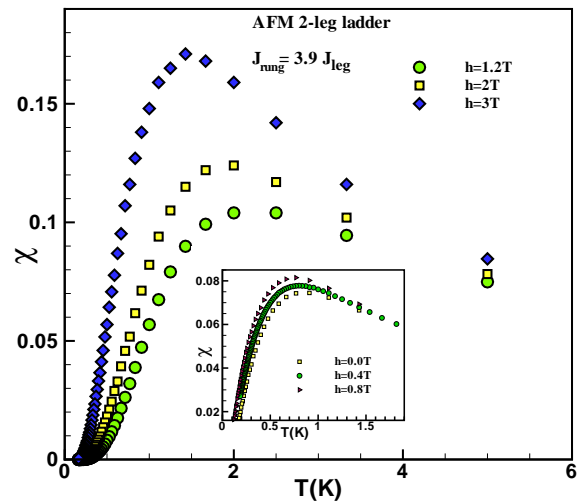,width=3.0in}}
 \caption{Magnetic susceptibility $\chi(T)$ versus temperature of isolated two-leg ladder
 at various magnetic field $h$. The inset shows $\chi(T)$ at low magnetic field. SSE QMC calculation carried out for Heisenberg model for $J_{leg}$=3.3 K and $J_{rung}$=12.9 K. There is a cross over from Curie-Weiss  law to an exponential behavior $\chi(T)\sim$exp$(-\Delta/T)/\sqrt{T}$ in the susceptibility curve indicating the spin gap in the systems.}\label{fig2}
\end{figure}

\subsection{Isolated two-leg ladders}

In Fig.~\ref{fig2}, the temperature dependence of the magnetic susceptibility
 is shown for isolated spin-1/2 two-leg ladder with the size of 2$\times$20 spins at different values of the magnetic field $h$, with temperature range \emph{T}=0.1 K to 7 K.
 In the inset of Fig.~\ref{fig2}, the low-temperature regime of the susceptibility is shown at three low magnetic fields.
 The exchange coupling ratio, $J_{rung}/J_{leg}=3.9$ is chosen according to evaluating of exchange interaction for (C$_{5}$H$_{12}$N)$_{2}$CuBr$_{4}$
or Cu$_{2}$(C$_{5}$H$_{12}$N$_{2}$)$_{2}$Cl$_{4}$.
 The reported results  here are qualitatively agreement  with  several
theoretical and experimental analyses of the two-leg ladder cuprate compounds susceptibility like
SrCu$_{2}$O$_{3}$ \cite{Azuma}, (C$_{5}$H$_{12}$N)$_{2}$CuBr$_{4}$ \cite{Watson} and vanadate compound MgV$_{2}$O$_{5}$ \cite{Millet}.
There is a cross over from Curie-Weiss  law to an exponential behavoir
$\chi(T)\sim$exp$(-\Delta/T)/\sqrt{T}$ in the susceptibility curve indicating the spin gap in these systems.
We have found the spin gap of decoupled two-leg ladders at different low magnetic fields using $\chi(T)\sim$exp$(-\Delta/T)/\sqrt{T}$.
A spin gap of $\Delta$=9 K at \emph{h}=0 is found from our QMC simulation of the magnetic susceptibility which is close to value about $\Delta=J_{rung}-J_{leg}$ determined by experimental results \cite{Ruegg,Watson}.
Spin gap decreases slightly to 7.4 K in the magnetic field of $h$=2 T.
For low magnetic fields,  the susceptibility
    goes up first with lowering temperature until it reaches to a maximum, then
    decreases down to zero at low temperature. A broad peak at $T$=2 K is
     interpreted as an existence of the gaped phase. In the high magnetic field,
      the susceptibility has no maximum.
      The low temperature behavior is qualitatively
      different from experimental results of Sr$_{14-x}$Ca$_{x}$Cu$_{24}$O$_{41}$ ladder superconductors due to existence of FM interaction in the chains\cite{Hiroi}.




 \begin{figure}[h]
\centerline{\psfig{file=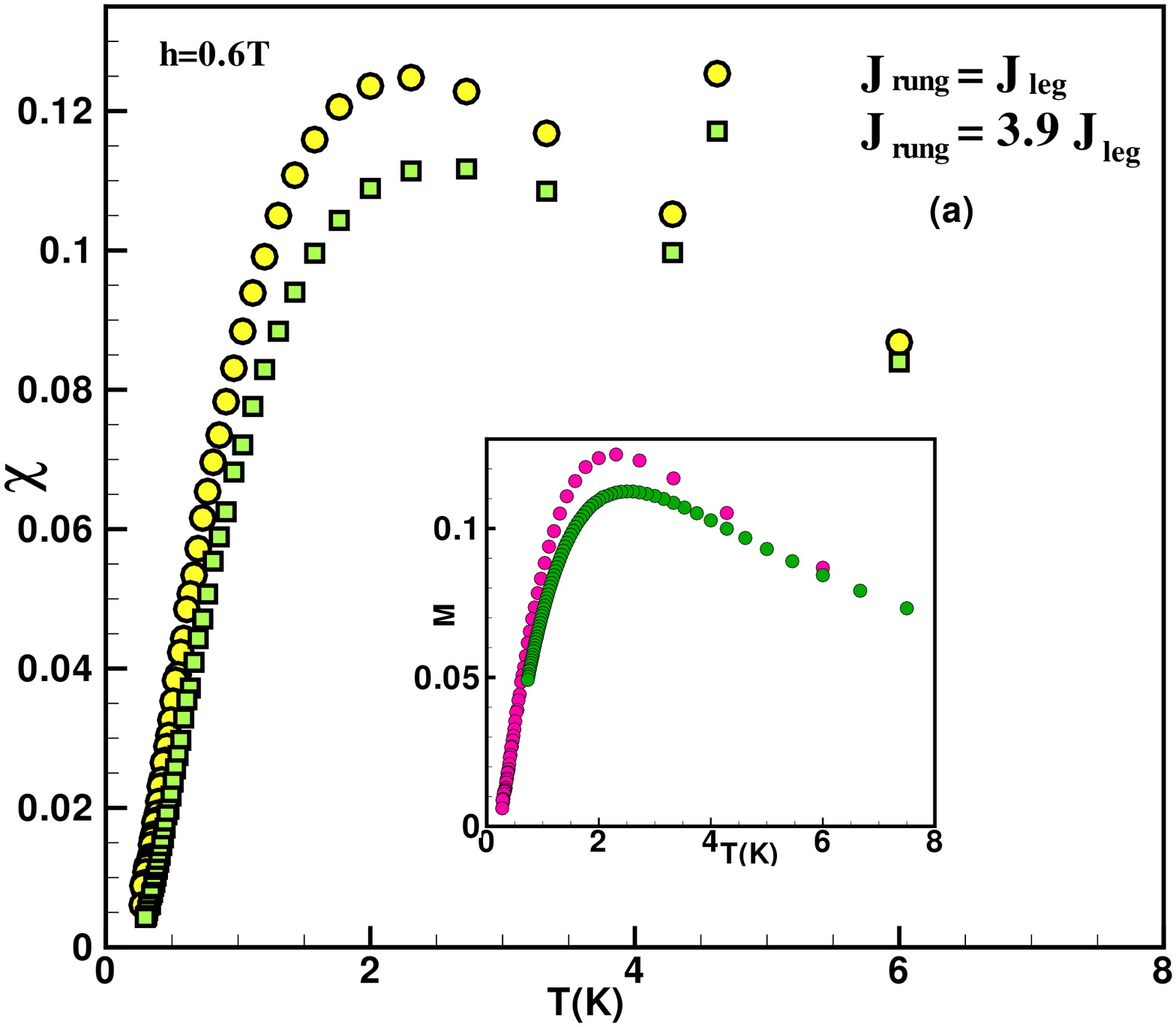,width=3.0in}}
\centerline{\psfig{file=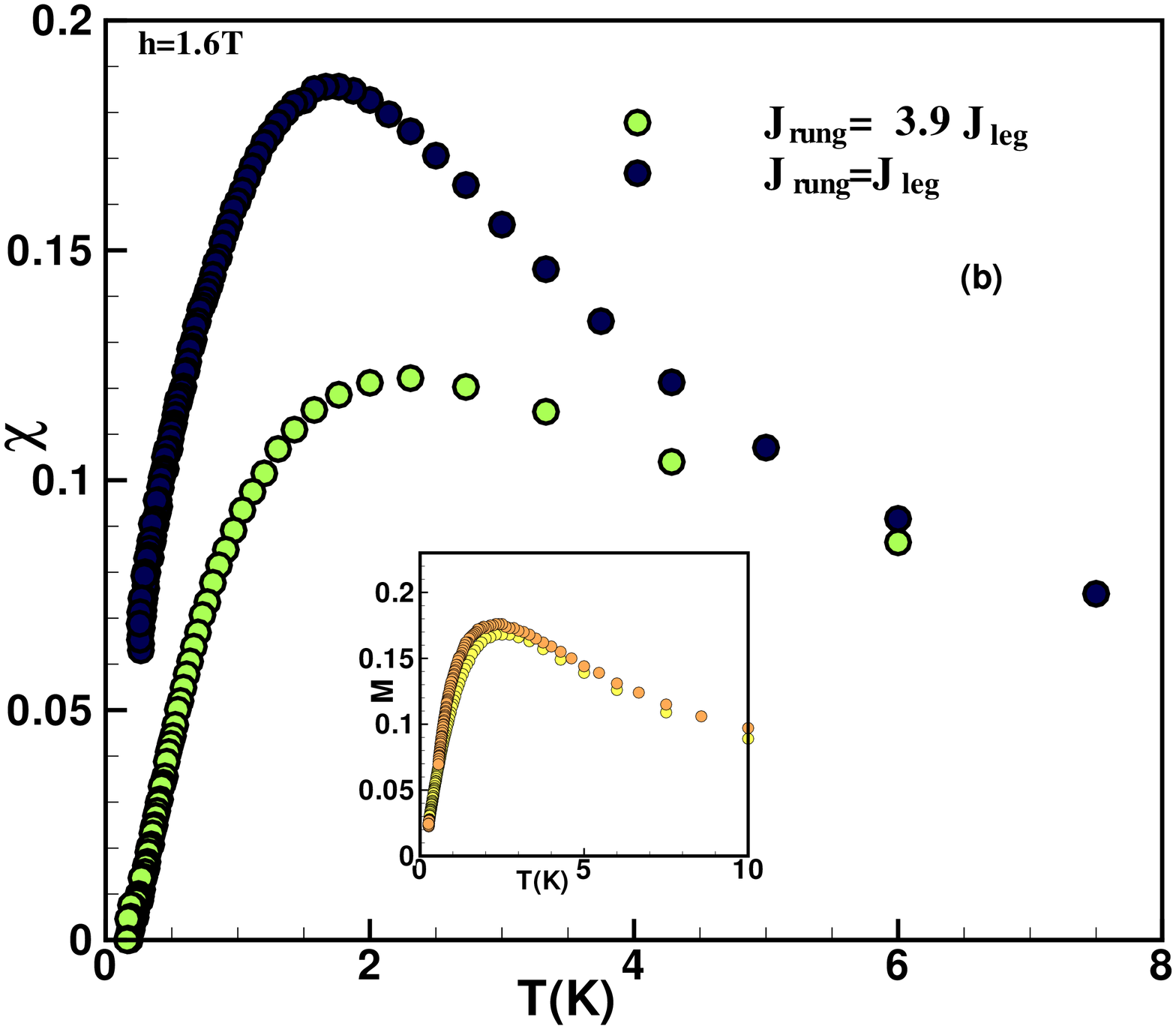,width=3.0in}}
 \caption{ Magnetic susceptibility $\chi(T)$ and Magnetization $M(T)$ (depicted in the inset) versus temperature of isolated two-leg ladder at (a) \emph{h}=0.6 T and (b) \emph{h}=1.6 T. SSE QMC calculation carried out for Heisenberg model for $J_{leg}$=3.3 K and $J_{rung}$=12.9 K.  Fig.~\ref{fig3}(b) shows that the spin ladder system with $J_{rung}/J_{leg}$=1 can be in a spin-liquid phase similar to the more familiar cases $J_{rung}/J_{leg}$=3.9}\label{fig3}
\end{figure}
Here to consider the effect of intra-ladder exchange interaction ratio $J_{rung}/J_{leg}$, we compare the magnetic properties of two systems with different  $J_{rung}/J_{leg}$=3.9  and $J_{rung}/J_{leg}$=1 at low magnetic fields.  As shown in Figs.~\ref{fig3}(a) and (b), the susceptibility is weakly  dependent on $J_{rung}/J_{leg}$ ratio at low magnetic fields. Susceptibility curve Fig.~\ref{fig3}(b) indicates that the spin singlet state strengthens and spin gap increases with enhancing the $J_{rung}/J_{leg}$ coupling ratio. Small differences can
be noticed in the magnetization curves (inset of Figs.~\ref{fig3}(a) and (b)). So, similar to the more familiar cases $J_{rung}/J_{leg}$=3.9, the results show that the spin ladder system with $J_{rung}/J_{leg}$=1 would be in a spin-liquid state.
As shown in Fig.~\ref{fig3}(a) in the range of weak magnetic field h=0.6T the spin gap  for $J_{rung}/J_{leg}$=1 is estimated about 7.5K, and for $J_{rung}/J_{leg}$=3.9 is estimated 8.5K.
In Fig.~\ref{fig3}(b), the spin gap with enhancing magnetic field to 1.6T  for two ratios of $J_{rung}/J_{leg}$ is estimated. The spin gap is estimated about 7.5K and 5K for $J_{rung}/J_{leg}$=3.9 and 1.0 respectively.


\begin{figure}[t]
\centerline{\psfig{file=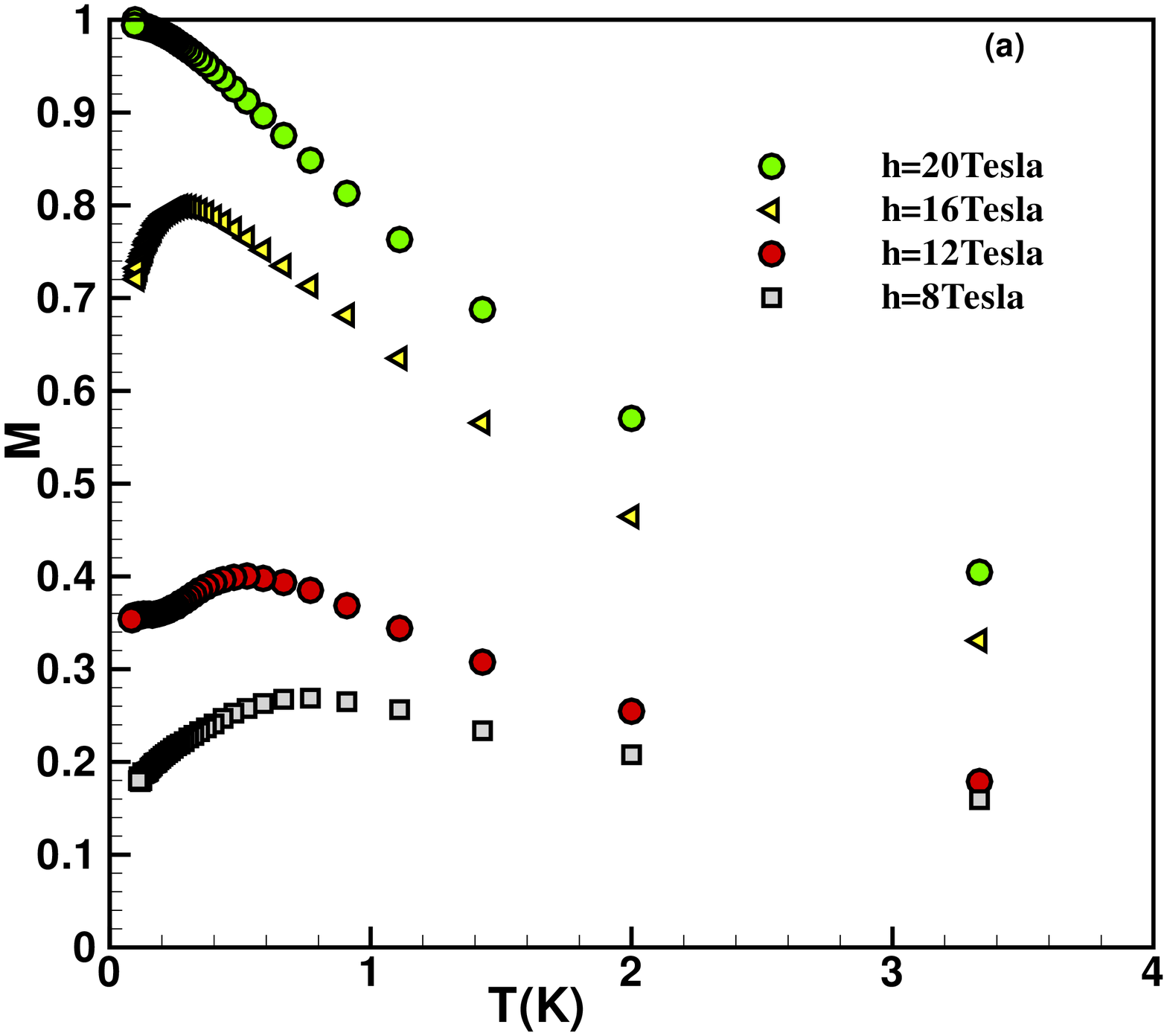,width=3.0in}}
\centerline{\psfig{file=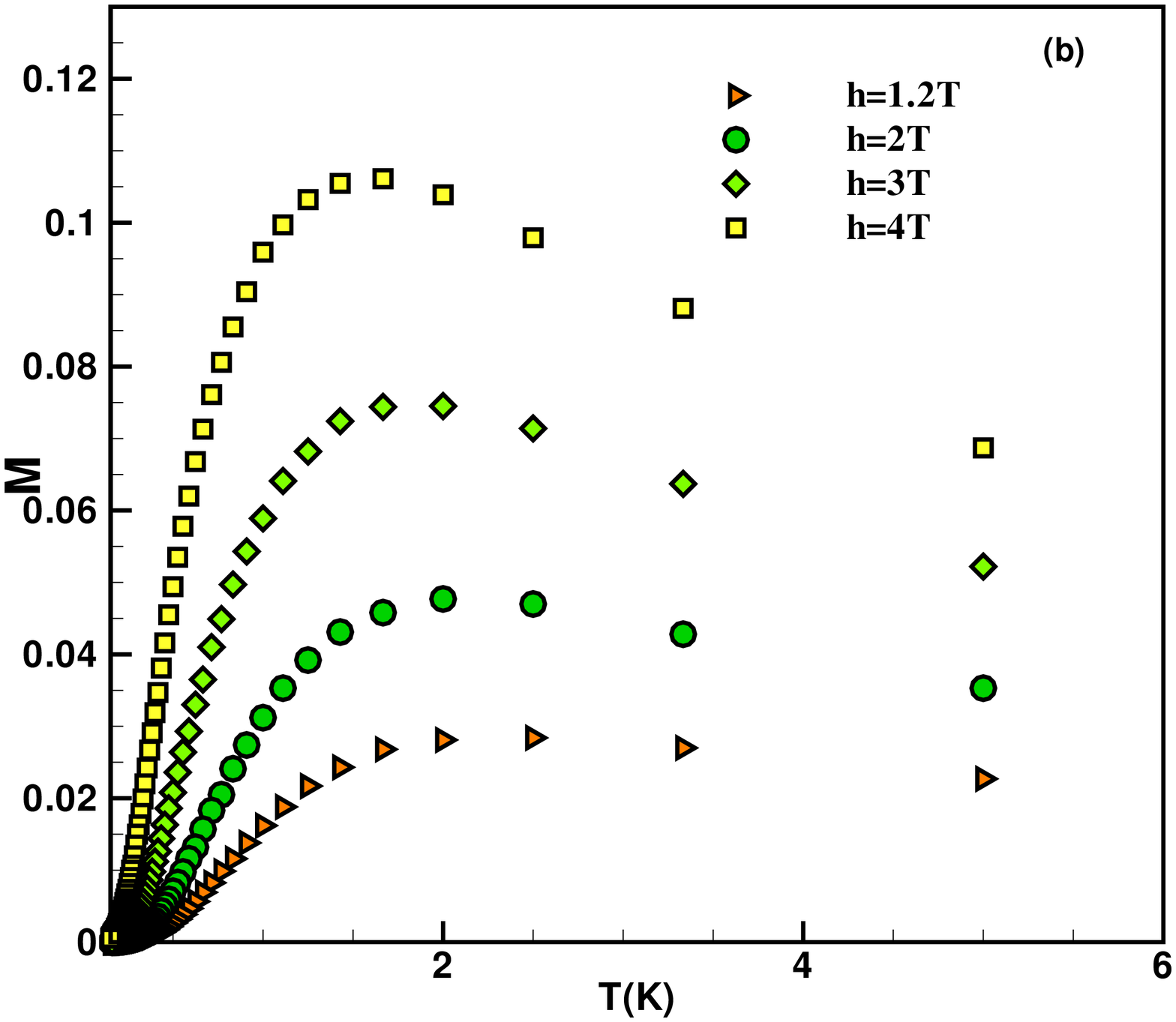,width=3.0in}}
 \caption{Magnetization $M(T)$ versus temperature of isolated two-leg ladder
 at various magnetic field $h$ (a) high magnetic fields (b) low magnetic fields. SSE QMC calculation carried out for Heisenberg model for $J_{leg}$=3.3 K and $J_{rung}$=12.9 K.}\label{fig4}
\end{figure}

Now, we also present the QMC results for the magnetization versus temperature $M(T)$ of decoupled spin-1/2 two-leg ladders at various magnetic fields in Fig.~\ref{fig4}. From these results, we determined the values of the critical fields as $h_{c1}\simeq8.5$T and $h_{c2}\simeq20$T.  When $h<h_{c1}$, magnetization first goes up at high temperature, then goes down to zero exponentially (see Fig.~\ref{fig4}(b)). For $h_{c1}<h<h_{c2}$, there are some maximum and minimum which separate the TLL phase from spin liquid and spin polarized phase respectively (Fig.~\ref{fig4}(a)). With lowering temperature, $M(T)$ begins to increase and saturates exponentially for $h>h_{c2}$.
\begin{figure}[t]
\centerline{\psfig{file=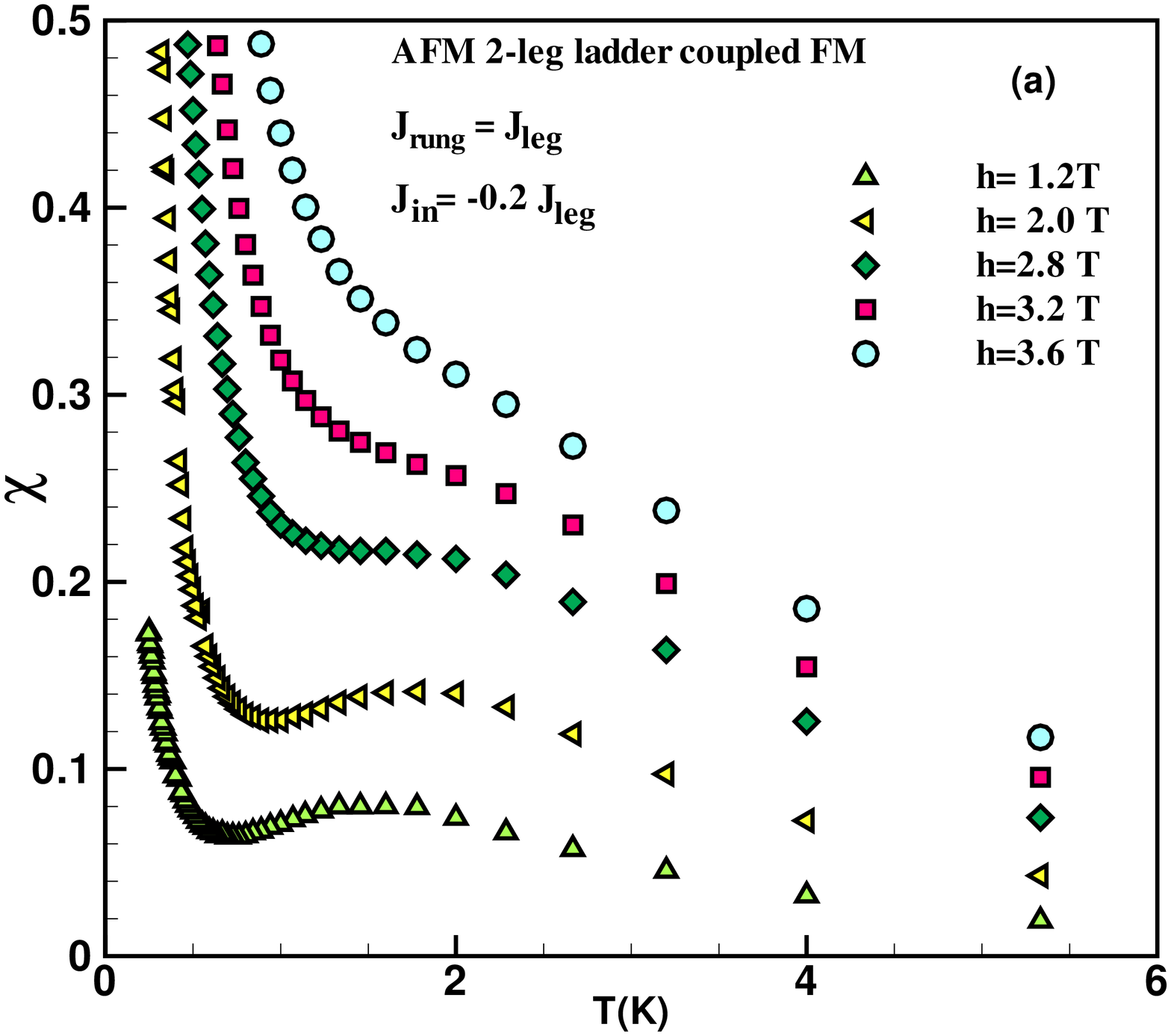,width=3.0in}}
\centerline{\psfig{file=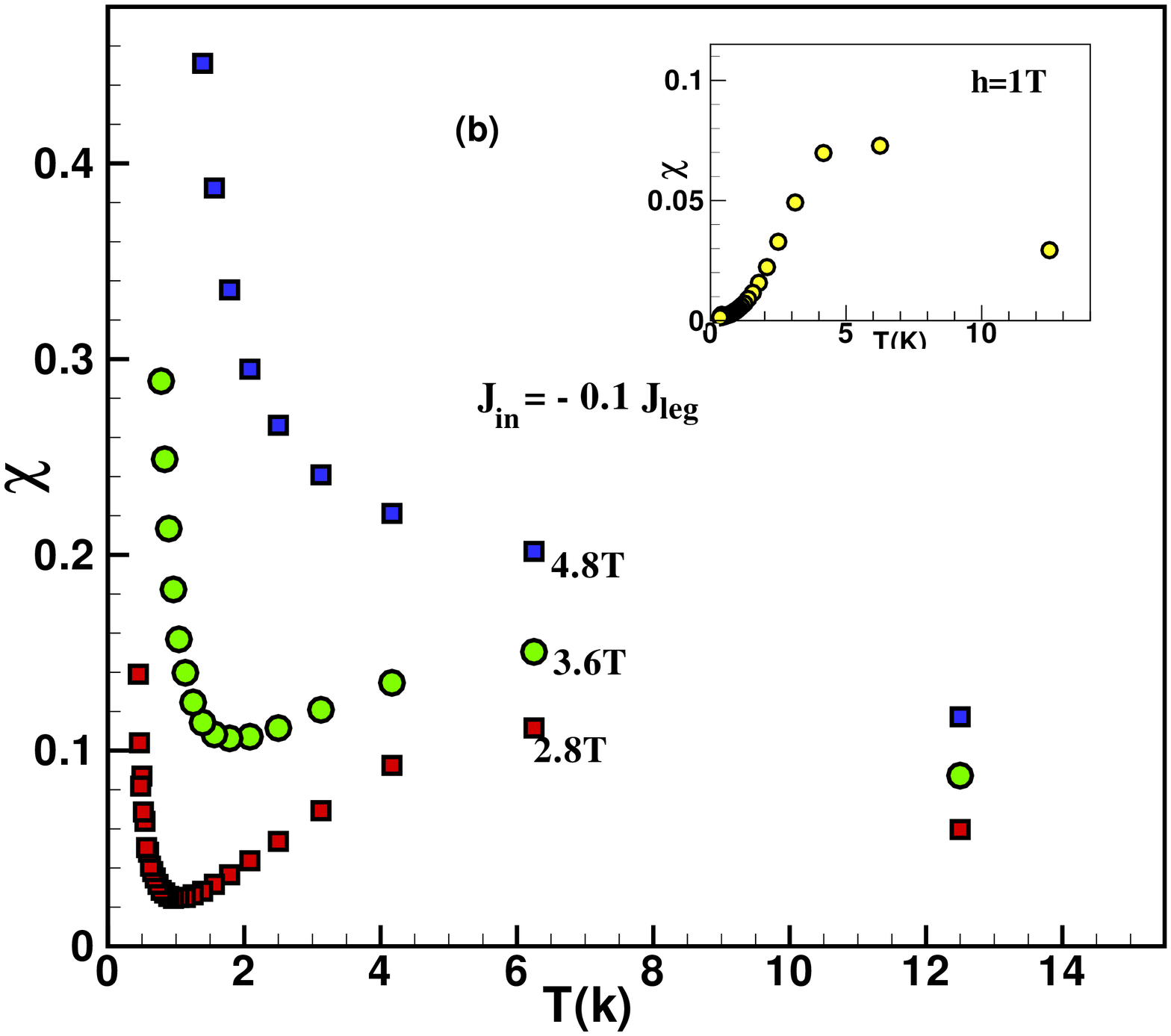,width=3.0in}}
 \caption{(a) Magnetic susceptibility $\chi(T)$ versus temperature of FM coupled two-leg ladder
 at various magnetic field $h$. SSE QMC calculation carried out for Heisenberg model for $J_{leg}$=3.9 K, $J_{rung}$=3.9 K, and $J_{in}$= -0.2$J_{leg}$. (b) Magnetic susceptibility $\chi(T)$ versus temperature of FM coupled two-leg ladders for inter-ladder interaction $J_{in}=-0.1J_{leg}$. SSE QMC calculation carried out for Heisenberg model for $J_{leg}$=3.9 K and $J_{rung}$=3.9 K.}\label{fig5b}
\end{figure}
\subsection{FM coupled two-leg ladders compared with isolated ladders}

\begin{figure}[t]
\centerline{\psfig{file=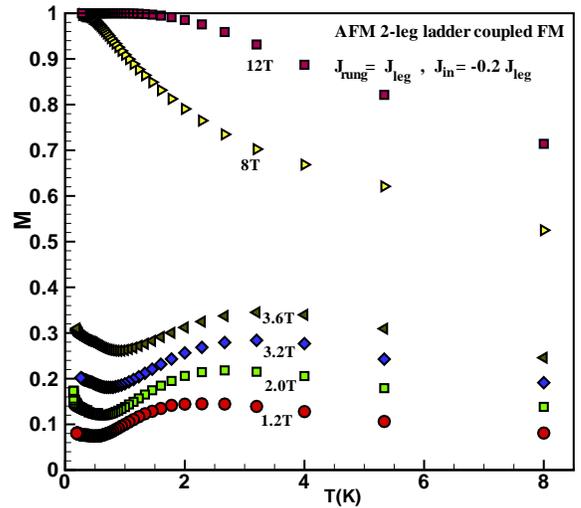,width=3.0in}}
 \caption{Magnetization $M(T)$ versus temperature of FM coupled two-leg ladder
 at various magnetic fields $h$. SSE QMC calculation carried out for Heisenberg model for $J_{leg}$=3.9 K, $J_{rung}$=3.9 K, and $J_{in}=-0.2J_{leg}$.}\label{fig6}
\end{figure}

Next, let us consider a case of such spin-1/2 two-leg ladders which are weakly coupled ferromagnetically with inter-ladder exchange interaction  $J_{in}=-0.1 J_{leg}$ and $J_{in}=-0.2 J_{leg}$. As we mentioned before, to better comparing of the results with isolated ladders we have performed QMC calculation in the low range of exchange coupling ($J_{leg}$=3.9 K). So,  temperature range of Curie-Weiss behavior and magnetic critical fields in our QMC simulation for the FM coupled ladders are one or two order of magnitude lower than experimental results of Sr$_{14-x}$Ca$_{x}$Cu$_{24}$O$_{41}$.
   Fig.~\ref{fig5b}(a) shows that temperature dependence of the susceptibility curve is found to be strongly dependent on magnetic fields.
  The susceptibility curves at temperature less than $T$=1 K have an upturn which is due to FM inter-ladder coupling. This behavior is absent in the magnetic susceptibility for decoupled two-leg ladders. The asymptotic behavior shows that the ground state of the system is in a magnetic order phase at $T=0$. In principle, by adding the inter-ladder interaction the ground state of the system undergoes a quantum phase transition to a magnetic order phase in magnetic fields larger than $h$=1.0 T. Qualitatively, there is a good similarity between
  the experimental results of  Sr$_{14-x}$Ca$_{x}$Cu$_{24}$O$_{41}$ and our QMC calculation by considering the inter-ladder interaction \cite{Hiroi}.
  But, the origin of the upturn behavior of the susceptibility is different from the experimental results. On the other hand,
   as shown in Fig.~\ref{fig5b}(a) the Curie-Weiss behavior due to inter-ladder interaction shifts to higher temperature with increasing the value of the magnetic field.
  As mentioned in the introduction, the Curie temperature about 20 K in the experimental results indicate the spin of Cu atoms contribution on the chains. The upturn temperature in Fig ~\ref{fig5b}(a) is one order of magnitude less than experimental results of  Sr$_{14-x}$Ca$_{x}$Cu$_{24}$O$_{41}$ due to considering of small $J_{rung}$ in our QMC simulation.
The magnetic field to make disappear the gaped phase ($h_{c1}$=1.0 T) is lower in the FM coupled two-leg ladder as compared with decoupled case. Therefore such a small value of $h_{c1}$
in FM coupled ladders places the physical properties of Sr$_{14-x}$Ca$_{x}$Cu$_{24}$O$_{41}$ in the vicinity
of the TLL phase.
The gaped phase (spin-singlet state) is not clear in the reported range of the magnetic field, $1.2 \leq h \leq 3.6$. Since, increasing magnetic field from $h_{c_1}=1.0 T$, breaks down the spin-singlet state and vanishes the gaped phase of the system. For the FM coupling less than $J_{in}=-0.2 J_{leg}$, we found a temperature dependence for the susceptibility
        similar to the case of $J_{in}=-0.2 J_{leg}$. In particular, as shown in Fig.~\ref{fig5b}(b), the weak inter-ladder interaction $J_{in}=-0.1 J_{leg}$ has the same qualitative effect on the
          thermodynamic properties within the range of magnetic fields $h_{c_1} \leq h \leq h_{c_2}$. In the inset of Fig.~\ref{fig5b}(b), we have plotted the magnetic susceptibility for a value of the magnetic filed less than the first quantum critical field, $h=1.0<h_{c_1}$. We found a temperature dependence for the susceptibility similar to previous results for the decoupled spin-1/2 two-leg ladder systems. In fact, the weak inter-ladder interaction $J_{in}=-0.1 J_{leg}$ has not remarkable effect on the thermodynamic properties at low  magnetic fields (less than the first quantum critical field)  due to the spin-singlet gaped nature of the individual ladders.
It is interesting to compare the evolution of critical magnetic field upon the considering of inter-ladder interaction.  For this purpose we have simulated $M(T)$ for FM coupled two-leg ladders $J_{in}= -0.2 J_{leg}$ of size
8$\times$20 spins at different values of the magnetic field \emph{h}= 1.2 T, 2.0 T, 2.8 T, 4.0 T, . . ., 12.0 T with temperature range \emph{T}= 0.1 K to 7 K. As shown in Fig.~\ref{fig6}, the value of  magnetic field to create the spin polarized state in FM coupled ladders is smaller than the magnetic field in the decoupled ladder case.
Also, the upturn feature at the low temperatures is absent
in the magnetization curves for the decoupled two-leg ladders.
At higher temperatures the effects of the weak FM inter-ladder interaction are not significant.
But, there is a departure of the decoupled two-leg ladders from
the FM coupled two-leg ladders at low temperature.
 The system exhibits extrema in $M(T)$ even in the low magnetic field due to FM interaction.

\begin{figure}[t]
\centerline{\psfig{file=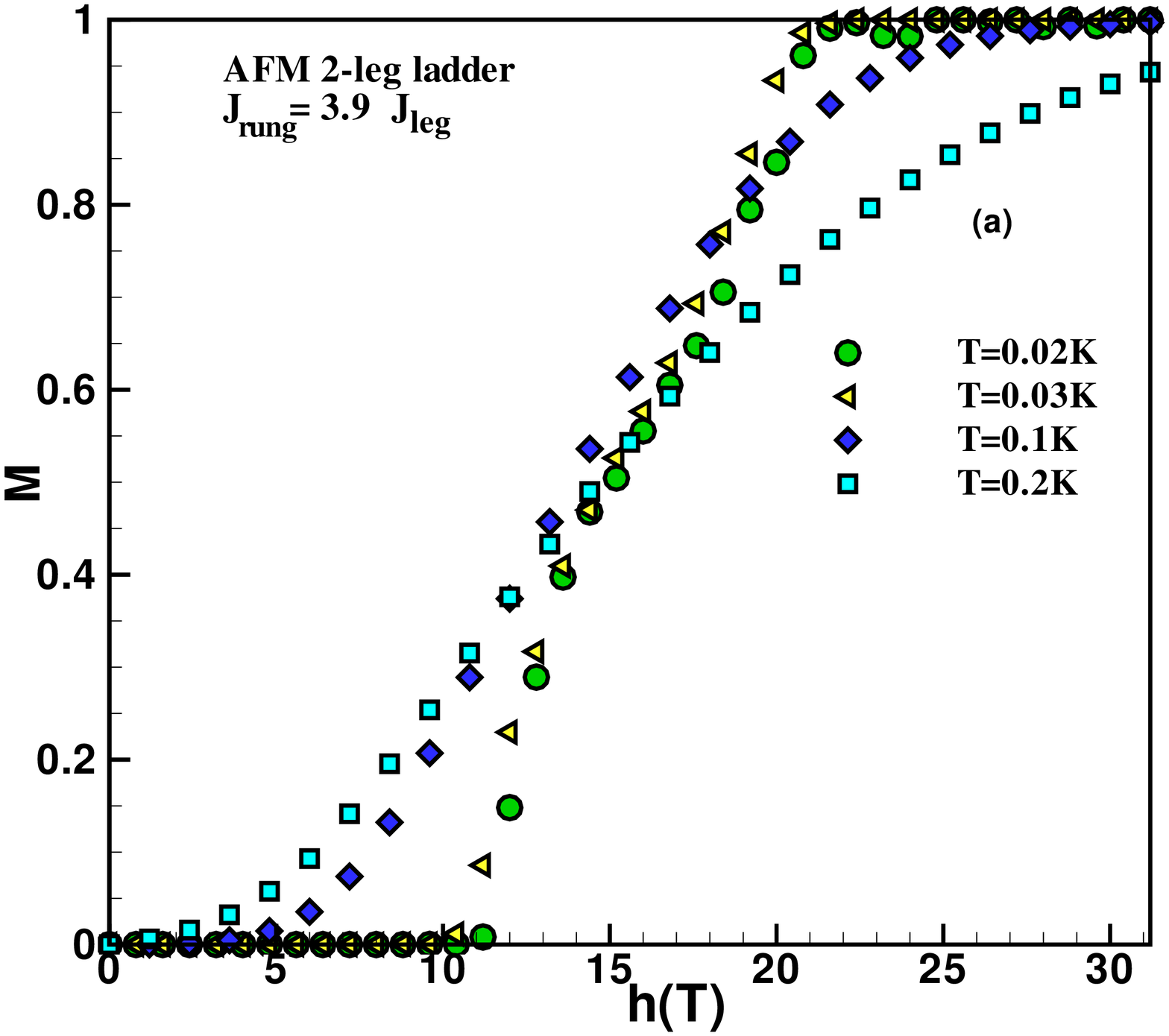,width=3.0in}}
\centerline{\psfig{file=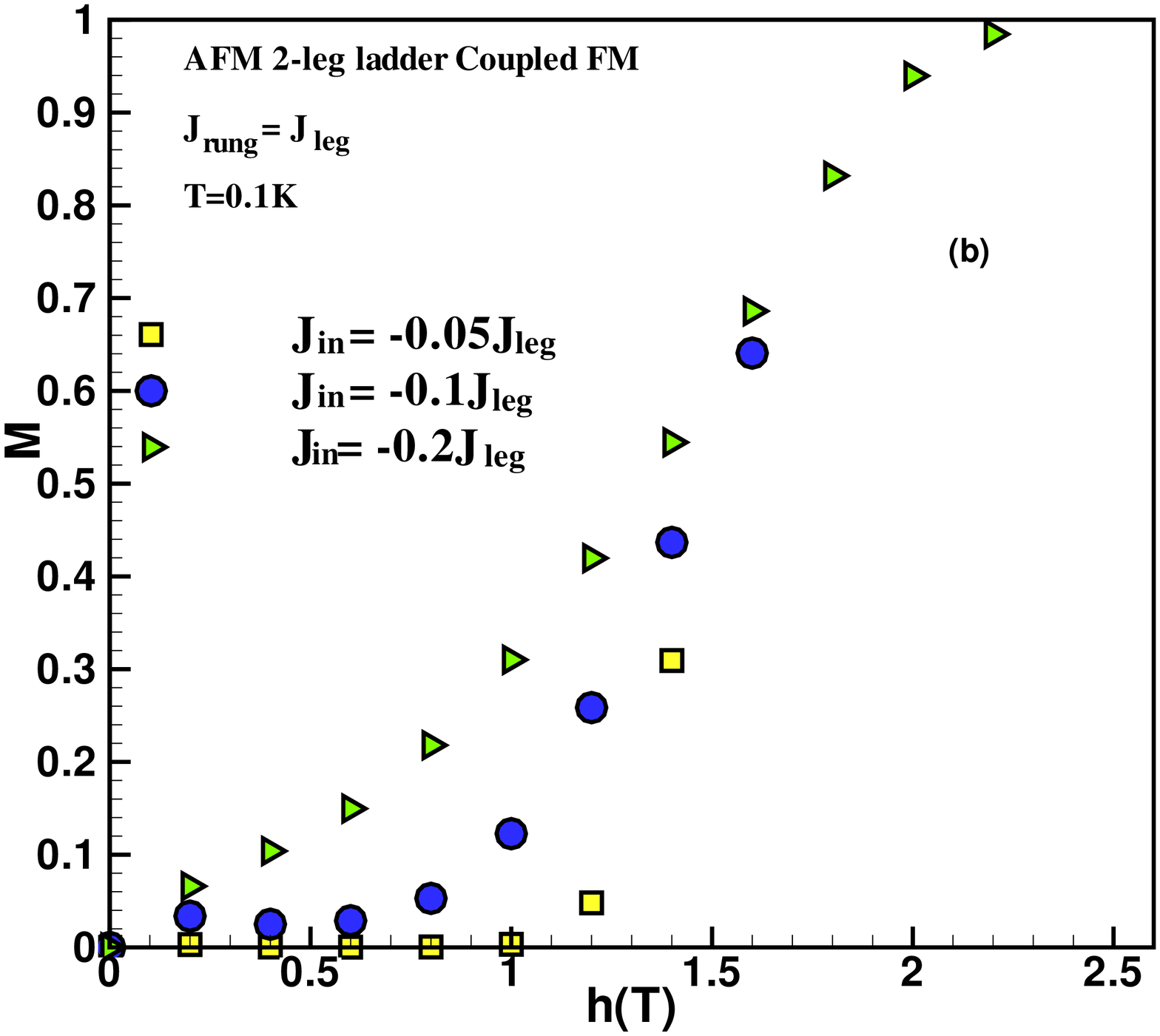,width=3.0in}}
 \caption{(a) Magnetization versus the magnetic field of isolated two-leg ladders at different temperature. SSE QMC calculation carried out for Heisenberg model for $J_{leg}$=3.3 K and $J_{rung}$=12.9 K. (b) Magnetization versus the magnetic field of FM coupled two-leg ladders at different $J_{in}$. SSE QMC calculation carried out for Heisenberg model for $J_{leg}$=3.9 K and $J_{rung}$=3.9 K.}\label{fig7}
\end{figure}

$M(h)$ is calculated for  spin 1/2 two-leg decoupled ladders of size
2$\times$20 with \emph{T}= 0.02 K, 0.03 K, 0.1 K, and 0.2 K with $J_{rung}/J_{leg}=3.9$ depicted in Fig.~\ref{fig7}(b).
 At low temperature below $T$= 0.1 K,  there is an energy gap in the magnetization
  versus the magnetic field due to the formation of gapped singlet
  state in spin-1/2 two-leg ladders.  The critical fields are approximated by our QMC calculation about 9.0 T and 20 T respectively, giving $h_{c1}=J_{rung}-J_{leg}$ and $h_{c2}=J_{rung}+2J_{leg}$.
  Fig.~\ref{fig7}(a) indicates that there is a gap in the system
even in the presence of FM inter-ladder interaction about $J_{in}=-0.1 J_{leg}$. The existence of the gap at
about  $J_{in}=-0.1 J_{leg}$ is consistent with susceptibility and magnetization curve depicted in Fig.~\ref{fig5b} (b). As expected, the spin gap reduces with increasing $J_{in}$.



Now, we also present the QMC results for the specific heat $C{_m}$ to find out the effect of
inter-ladder FM exchange interaction in the spin-1/2 two-leg ladder systems.
 We consider the effect of inter-ladder interaction with the ratio of $J_{in}/J_{leg}=0.2$ in the strong coupling limit i.e. $J_{rung}/J_{leg}=3.9$, and $J_{rung}/J_{leg}=1$, in order to see the changes of cross-over from one phase to the another phase. Specially, we consider this effect within the range of  $h_{c_1} \leq h \leq h_{c_2}$ (TLL phase). In Fig.~\ref{fig8}(a) we have plotted $C{_m}$ curve for an isolated two-leg ladder system  with the ratio of  $J_{rung}/J_{leg}=3.9$ and AFM two-leg ladders coupled FM with the same ratio. In both of them we observe a remarkable second peak at very low temperatures which is known as the indication of the existence of the  TLL phase. As expected, the inter-ladder interaction has not changed the behaviour of $C{_m}$. Next, we consider the effect of inter-ladder interaction with coupling ratio of $J_{rung}/J_{leg}=1$ within the range of TLL phase. For isolated two-leg ladder and even, AFM two-leg ladder coupled FM the second peak was observed in Fig.~\ref{fig8}(b), however it is hard to see the second peak for coupled ladders in this case due to the finite size effects.


 \begin{figure}[t]
\centerline{\psfig{file=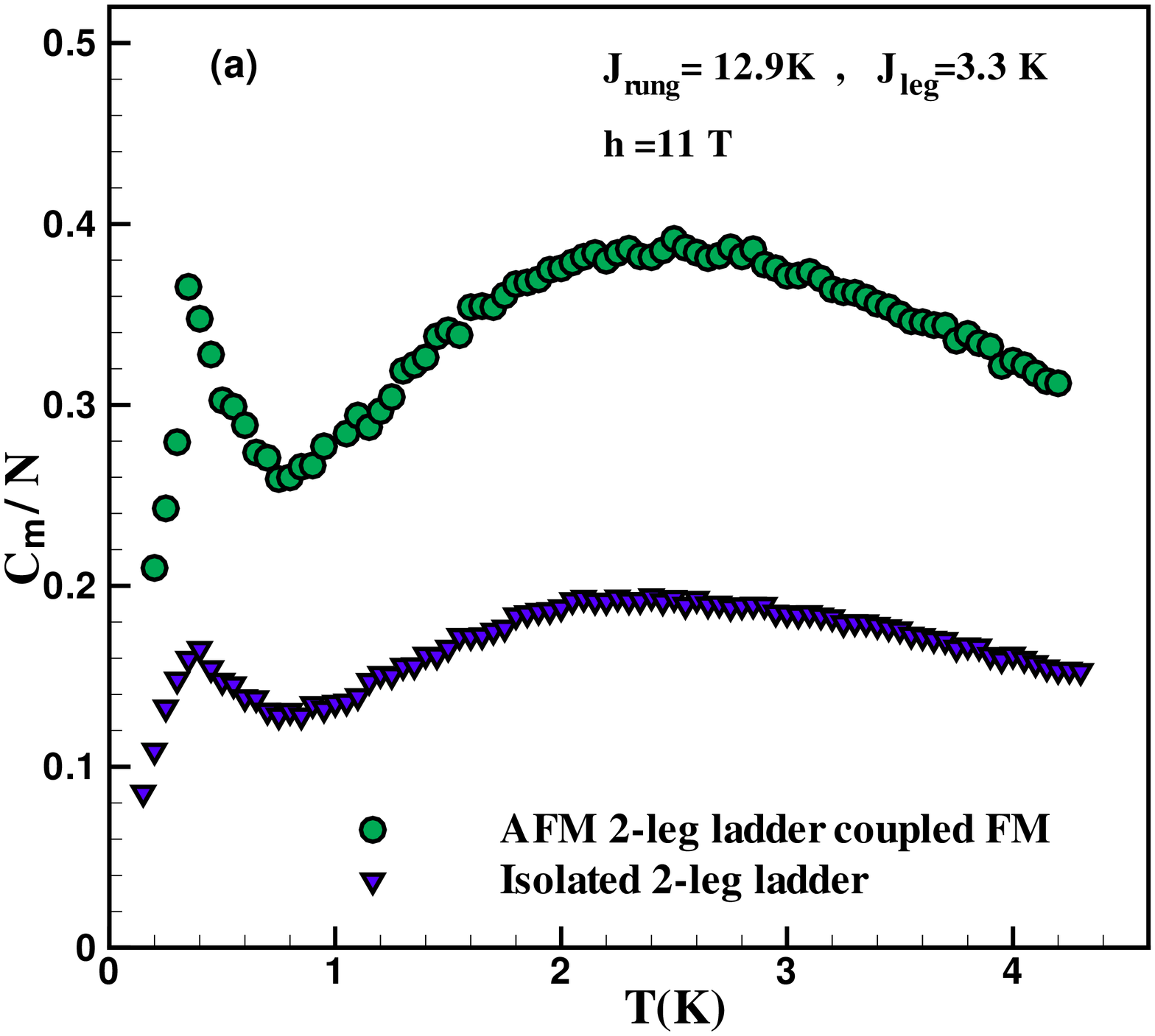,width=3.0in}}

\end{figure}
 \begin{figure}[t]
\centerline{\psfig{file=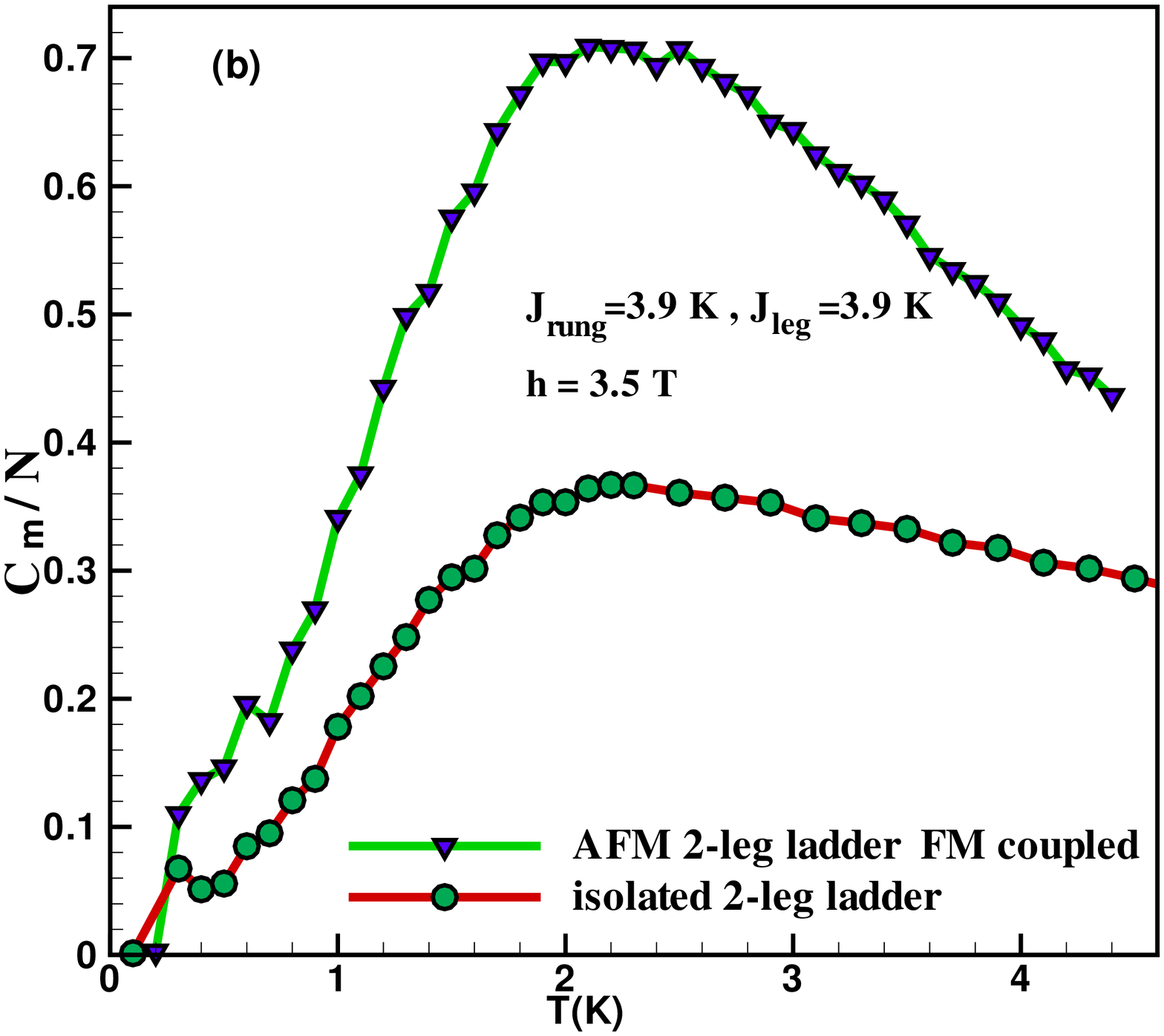,width=3.0in}}
 \caption{ Magnetic specific heat per cite versus temperature of both isolated two-leg ladders and FM coupled two-leg ladders at (a) intermediate field (TLL phase) for $J_{rung}$=3.9$J_{leg}$   (b) intermediate field (TLL phase) for $J_{rung}$=$J_{leg}$.  SSE QMC calculation carried out for Heisenberg model for $J_{in}=-0.2J_{leg}$}\label{fig8}
\end{figure}

\begin{figure}[t]
\centerline{\psfig{file=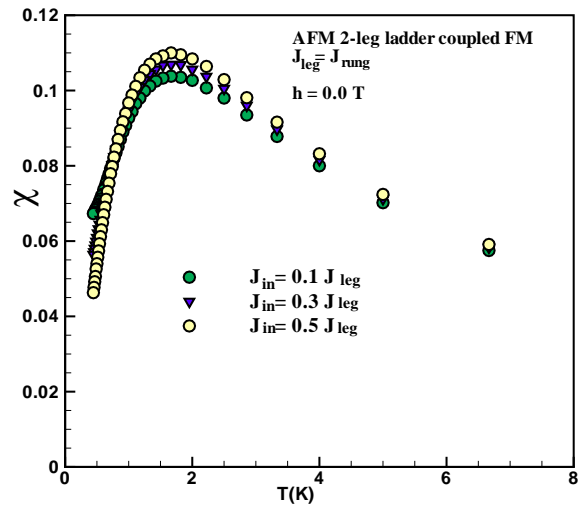,width=3.0in}}
 \caption{Magnetic susceptibility versus temperature of FM coupled two-leg ladders at different inter-ladder exchange interaction. To compare the existence of gapped singlet state, we have shown the curve for three $J_{in}$. SSE QMC calculation carried out for Heisenberg model for $J_{leg}$=3.9 K and $J_{rung}$=3.9 K.}\label{fig9}
\end{figure}

Let us see what occurs when we attempt to include the inter-ladder coupling effects in the absence of magnetic fields. So, we have carried out the simulation QMC for different $J_{in}/J_{leg}$ in the zero magnetic field Fig.~\ref{fig9}. Existence of gapped phase has been found in the magnetic susceptibility curve up to $J_{in}/J_{leg}= -0.5$.
   Our results are agreement to the QMC simulations of Miyahara et al. \cite{Miyahara} on the FM coupled two-leg ladder (trellis layer) system. So, in the absence of magnetic field,
temperature-dependent thermodynamic behavior of spin-1/2 two-leg ladder is similar to
   coupled one. The spin gap slightly decreases from $\Delta$=2 K to $\Delta$=1.6 K with increasing $J_{in}/J_{leg}$ up to $-0.5$.

 In the presence of magnetic fields, as shown in Fig.~\ref{fig2} and Fig.~\ref{fig5b}(a) or Fig.~\ref{fig5b}(b),
  the $\chi(T)$ temperature-dependent behavior of susceptibility of spin-1/2 two-leg ladder is quite different
   from coupled one, suggesting the inter-ladder exchange interaction needs to
   account for spin ladder systems like vanadate compound MgV$_{2}$O$_{5}$, and cuprate superconductor Sr$_{14-x}$Ca$_{x}$Cu$_{24}$O$_{41}$.
   As shown in Fig.~\ref{fig5b}(a) or Fig.~\ref{fig5b}(b), the magnetic susceptibility of coupled two-leg ladders is strongly temperature-dependent below $T$=2 K.
 The upturn behavior appears in the QMC simulation of susceptibility at low temperature in
 the magnetic fields. As we have mentioned in the previous part, the asymptotic behavior confirms the existence of the magnetic order at $T=0$.

\section{Conclusion}\label{sec3}

IIn summary, we have calculated the thermodynamic properties of
(C$_{5}$H$_{12}$N)$_{2}$CuBr$_{4}$ and Sr$_{14-x}$Ca$_{x}$Cu$_{24}$O$_{41}$ crystal as a spin-1/2 AFM two-leg ladder. We have performed stochastic series
expansion QMC to investigate the effect of the FM inter-ladder exchange
interaction on the low-temperature behavior of the system by considering the  AFM coupled ladders.
A remarkable upturn behavior of susceptibility is observed at low temperature in the low magnetic fields and the Curie-Weiss behavior due to inter-ladder interaction shifts to higher temperature with increasing magnetic fields.
In the absence of magnetic field, temperature dependence thermodynamic behavior of spin-1/2 two-leg ladder is similar to coupled one up to $-0.5 J_{leg}$ interaction. But, the gaped phase is not clear in the $J_{in}=-0.2 J_{leg}$ magnetic susceptibility of FM coupled two-leg ladder even at low magnetic fields.
 Although, in the absence of magnetic field,
the thermodynamic behavior of spin-1/2 two-leg ladder is similar to
   coupled one, the magnetic field to make disappear the gaped phase is lower in the FM coupled two-leg ladder as compared with decoupled case.  In the case of FM coupled ladders one would reach quite a large sensitivity of $\Delta$ to the magnetic field.
Although for the case of Sr$_{14-x}$Ca$_{x}$Cu$_{24}$O$_{41}$, the compound consists of weakly FM coupled ladders about $J_{in}=-0.1 J_{leg}$, but such enhancement of coupled interaction may occur upon chemical substitution.

\vspace{0.3cm}


\end{document}